\documentstyle{cupconf}
\begin{document}
%
%
%
\def\abs#1{\left| #1 \right|}
\def\EE#1{\times 10^{#1}}
\def\gcm{\rm ~g~cm^{-3}}
\def\cm3{\rm ~cm^{-3}}
\def\kms{\rm km~s^{-1}}
\def\cms{\rm ~cm~s^{-1}}
\def\ergs{\rm ~erg~s^{-1}}
\def\ergscm{\rm ~erg~s^{-1}~cm^{-2}}
\def\isotope#1#2{\hbox{${}^{#1}\rm#2$}}
\def\wl{~\lambda}
\def\wll{~\lambda\lambda}
\def\HI{{\rm H\,I}}
\def\HII{{\rm H\,II}}
\def\HeI{{\rm He\,I}}
\def\HeII{{\rm He\,II}}
\def\HeIII{{\rm He\,III}}
\def\CI{{\rm C\,I}}
\def\CII{{\rm C\,II}}
\def\CIII{{\rm C\,III}}
\def\CIV{{\rm C\,IV}}
\def\NI{{\rm N\,I}}
\def\NII{{\rm N\,II}}
\def\NIII{{\rm N\,III}}
\def\NIV{{\rm N\,IV}}
\def\NV{{\rm N\,V}}
\def\NVI{{\rm N\,VI}}
\def\NVII{{\rm N\,VII}}
\def\OI{{\rm O\,I}}
\def\OII{{\rm O\,II}}
\def\OIII{{\rm O\,III}}
\def\OIV{{\rm O\,IV}}
\def\OV{{\rm O\,V}}
\def\OVI{{\rm O\,VI}}
\def\OVII{{\rm O\,VII}}
\def\OVIII{{\rm O\,VIII}}
\def\CaI{{\rm Ca\,I}}
\def\CaII{{\rm Ca\,II}}
\def\NeI{{\rm Ne\,I}}
\def\NeII{{\rm Ne\,II}}
\def\NeIII{{\rm Ne\,III}}
\def\NeIV{{\rm Ne\,IV}}
\def\NeV{{\rm Ne\,V}}
\def\NaI{{\rm Na\,I}}
\def\NaII{{\rm Na\,II}}
\def\NiI{{\rm Ni\,I}}
\def\NiII{{\rm Ni\,II}}
\def\FeI{{\rm Fe\,I}}
\def\FeII{{\rm Fe\,II}}
\def\FeIII{{\rm Fe\,III}}
\def\FeV{{\rm Fe\,V}}
\def\FeVII{{\rm Fe\,VII}}
\def\CoII{{\rm Co\,II}}
\def\CoIII{{\rm Co\,III}}
\def\ArI{{\rm Ar\,I}}
\def\MgI{{\rm Mg\,I}}
\def\MgII{{\rm Mg\,II}}
\def\SiI{{\rm Si\,I}}
\def\SiII{{\rm Si\,II}}
\def\SiIII{{\rm Si\,III}}
\def\SiIV{{\rm Si\,IV}}
\def\SiVI{{\rm Si\,VI}}
\def\SI{{\rm S\,I}}
\def\SII{{\rm S\,II}}
\def\SIII{{\rm S\,III}}
\def\SIV{{\rm S\,IV}}
\def\SVI{{\rm S\,VI}}
\def\FeI{{\rm Fe\,I}}
\def\FeII{{\rm Fe\,II}}
\def\FeIII{{\rm Fe\,III}}
\def\FeIV{{\rm Fe\,IV}}
\def\FeVII{{\rm Fe\,VII}}
\def\kI{{\rm k\,I}}
\def\kII{{\rm k\,II}}
\def\La{{\rm Ly}\alpha}
\def\Ha{{\rm H}\alpha}
\def\Hb{{\rm H}\beta}
\def\Lya{{\rm Ly}\alpha}
\def\etscale#1{e^{-t/#1^{\rm d}}}
\def\etscaleyr#1{e^{-t/#1\,{\rm yr}}}
\def\sigmaKN{\sigma_{\rm KN}}
\def\ncrit{n_{\rm crit}}
\def\Emax{E_{\rm max}}
\def\chieff{\chi_{\rm eff}^{\phantom{0}}}
\def\chieffi{\chi_{{\rm eff},i}^{\phantom{0}}}
\def\chiion{\chi_{\rm ion}^{\phantom{0}}}
\def\chiioni{\chi_{{\rm ion},i}^{\phantom{0}}}
\def\Gammaion{\Gamma_{\!\rm ion}}
\def\Mcore{M_{\rm core}}
\def\Rcore{R_{\rm core}}
\def\Vcore{V_{\rm core}}
\def\Menv{M_{\rm env}}
\def\Venv{V_{\rm env}}
\def\Vej{V_{\rm ej}}
\def\Vcthou{\left( {\Vcore \over 2000 \rm \,km\,s^{-1}} \right)}
\def\KK{\rm ~K}
\def\Msun{~M_\odot}
\def\Msunyr{~M_\odot~{\rm yr}^{-1}}
\def\Mdot{\dot M}
\def\uten{u_{\rm 10}}
\def\tyr{t_{\rm yr}}
\def\gff{g_{\rm ff}}
\def\Tex{T_{\rm ex}}
\def\lsim{\!\!\!\phantom{\le}\smash{\buildrel{}\over
  {\lower2.5dd\hbox{$\buildrel{\lower2dd\hbox{$\displaystyle<$}}\over
                               \sim$}}}\,\,}
\def\gsim{\!\!\!\phantom{\ge}\smash{\buildrel{}\over
  {\lower2.5dd\hbox{$\buildrel{\lower2dd\hbox{$\displaystyle>$}}\over
                               \sim$}}}\,\,}
\input{psfig}
\input epsf
\title[Circumstellar interaction in Type Ia supernovae]{Supernova progenitor 
constraints from circumstellar interaction: Type~Ia}

\author[Lundqvist \& Cumming]%
{P\ls E\ls T\ls E\ls R\ns L\ls U\ls N\ls D\ls Q\ls V\ls I\ls S\ls T\ns\\
\and 
\ns R\ls O\ls B\ls E\ls R\ls T\ns J.\ns C\ls U\ls M\ls M\ls I\ls N\ls G}
\affiliation{Stockholm Observatory, S-133~36 Saltsj\"obaden, Sweden\\ 
(e-mail peter@astro.su.se, robert@astro.su.se)\\[\affilskip]}

\setcounter{page}{1}

\maketitle

\begin{abstract}
Searching for the presence of a circumstellar medium (CSM) is a direct
observational way to discriminate between different types of
progenitor systems for Type Ia supernovae (SNe Ia). We have modeled
whether such gas may give rise to detectable emission, especially in
H$\alpha$, and compare the models with observations of SN 1994D. We
obtain $\Mdot \la 2.5 \times 10^{-5} \Msunyr$ for a wind speed of
$10~\kms$. We find that X-ray observations in the range $5 - 10$~keV,
e.g., with {\it AXAF}, provide the most useful limits on the mass
loss, while high-resolution optical spectroscopy offers the only
direct way of identifying circumstellar hydrogen.
\end{abstract}


\section{Introduction}
SNe Ia are thought to be exploding white dwarfs in binary systems. The
most likely type of progenitor system (Branch et al.\ 1995; see also Iben \&
Tutukov 1984) is a C-O
white dwarf accreting H/He-rich gas from a companion, either from its
wind or through Roche lobe overflow. Coalescing pairs of C-O white
dwarfs are also possible, while accreting sub-Chandrasekhar-mass white
dwarfs are less likely (Branch et al.\ 1995). In the
non-coalescing scenarios, circumstellar gas will be present.  The
composition and geometry of this depend on the type of progenitor
system.  If the CSM emits detectable radiation, or absorbs radiation
from the supernova, this can be used to distinguish between types of
progenitor system. For example, any circumstellar lines of hydrogen
would have to come from gas lost by the companion, and the luminosity
of these lines are therefore particularly sensitive to the type of
system.

The interaction of the ejecta with the putative CSM can generate radio
(Boffi \& Branch 1995) and X-ray emission (Schlegel \& Petre 1993). If
dusty and asymmetric, the CSM may also result in polarization of the
supernova light (Wang et al. 1996). None of these studies have
resulted in a detection.  A different approach was taken by us in
Cumming et al.\ (1996; henceforth CLSPK96).  We used a high-resolution
optical spectrum of SN 1994D around H$\alpha$ taken only 6.5 days
after the explosion to search for circumstellar hydrogen. The
observations were compared with detailed photoionization calculations
to establish a limit on the mass loss from the progenitor system.  In
addition, we discussed the effect of asymmetry of the CSM, and
compared the sensitivity of optical studies to those at other
wavelengths. Here we expand this discussion.  In particular, we check
the sensitivity of our results to the adopted maximum velocity of the
ejecta, and demonstrate that absorption of soft ($\la$ 1$-$2~keV)
X-rays by the CSM must be taken into account in interpreting X-ray
limits.  Following from this we reassess the X-ray limit from SN 1992A
reported by Schlegel \& Petre (1993).

\section{Circumstellar excitation and expected circumstellar emission}

There are four sources of radiation which could excite the CSM of SN
Ia (CLSPK96): the radiation accompanying the supernova shock breakout,
$\gamma$-rays from the decay of $^{56}$Ni, the radiation emitted by
the progenitor prior to explosion (which can `preionize' the CSM), and
radiation from the interaction of the ejecta with the CSM itself.  Of
these, the first two are not important at all, while preionization is
only observationally important for SN Ia in the Magellanic Clouds, or
closer (CLSPK96).  For more distant supernovae, the only important
excitation mechanism is the radiation from the circumstellar
interaction.
\begin{figure}
\vspace{10cm}
\includegraphics{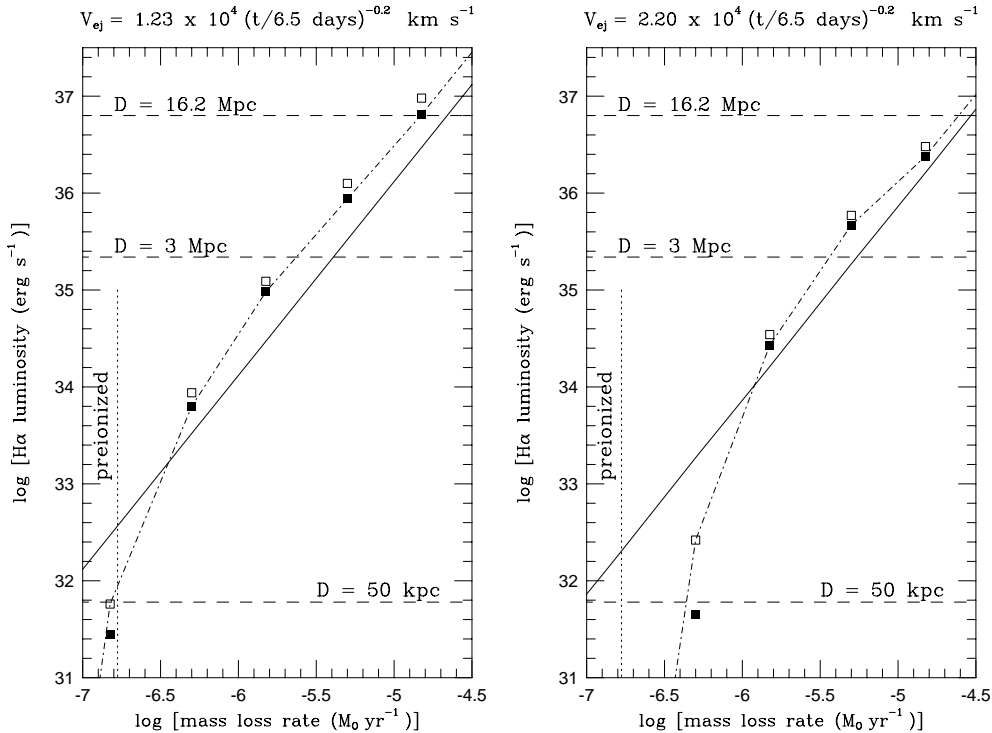}
\caption[]{Luminosity of circumstellar H$\alpha$ from a Type Ia
supernova at 6.5 days after explosion, as a function of mass loss from
the progenitor system, assuming a wind velocity of $10 \kms$.  Squares
show model calculations where the wind is ionized by the radiation
from the region of circumstellar interaction. The solid lines show the
corresponding luminosities for a fully ionized wind at
$2\times10^4$~K.  The maximum velocity of the ejecta is given at the
top of each panel.  Preionization by the progenitor white dwarf is
only important for $\Mdot / \uten \la 2 \times 10^{-7} \Msunyr$. See
text for details.}
\end{figure}
In Figure 1, we present the calculated circumstellar H$\alpha$ 
luminosity at 6.5 days as a function of mass
loss from the progenitor system. We show results both for $\Vej
\approx 1.23 \times 10^9~\kms$ and $\Vej \approx 2.20 \times
10^9~\kms$ at 6.5 days, where $\Vej$ is the maximum velocity of the ejecta.
For a wind with density falling off as $\rho \propto r^{-2}$, 
the interaction model of Chevalier (1982) predicts that the
ejecta are slowed down at a rate $\propto t^{-1/(n-2)}$. Here $n$ is
the power law index for the density of the unshocked ejecta,
$\rho_{\rm ej} \propto r^{-n}$. We use $n = 7$, which is a reasonable
approximation to the commonly used W7 model of Nomoto, Yoki, \&
Thielemann (1984), resulting in $\Vej \propto t^{-0.2}$.
For the CSM we assume solar abundances.

The ionizing radiation in the interaction model comes from the
C/O-rich ejecta shocked by the reverse shock propagating
(in mass coordinate) into the supernova.  The temperature of this gas,
$T_{\rm rev}$, at 6.5 days increases with the velocity of the ejecta,
and is $\sim 1 \times 10^8$~K and $\sim 4 \times 10^8$~K
for the two velocities shown in Figure 1.  
The spectrum of the ionizing radiation is therefore a free-free
spectrum with $kT_{\rm rev} \sim 10$~keV and $kT_{\rm rev} \sim
40$~keV, respectively. However, for low values of $\Mdot$, electrons
and ions in the shocked ejecta are not in energy equipartition, which
affects the level and cutoff energy of the ionizing flux. Figure 1
shows the resulting H$\alpha$ emission from the photoionized wind for
two cases: full equipartition between electrons and ions, and an
electron temperature which is a factor of 2 below the equipartition
value. The dashed-dotted lines join the most likely models.
A general feature of these models
is that the circumstellar shock is preceded by an ionization precursor. 
The thickness of this photoionized region increases with $\Mdot / u$
(CLSPK96), but decreases with $\Vej$. The reason for the latter is
mainly that the ratio of the dynamical time scale of the shock to the
ionization time scale of the CSM decreases with increasing $\Vej$.
Using the H$\alpha$ limit for SN 1994D (distance `16.2 Mpc' in Figure
1), and the fact that $\Vej$ of the supernova at 6.5 days may have
been as fast (Patat et al.\ 1996) as in the fastest model
in Figure 1, we obtain $\Mdot / \uten \la 2.5 \times 10^{-5} \Msunyr$. 
Here $\uten$ is the wind speed in $10 \kms$. This limit is a 
factor $\sim 1.7$ higher than in CLSPK96, where we assumed somewhat slower 
ejecta, and close to the upper limit observed for mass loss rates in
symbiotic systems. Observing SNe Ia in the Local Group (limit marked
`3 Mpc' in Figure 1) for 10\,000 s as early as 3 days after explosion,
should take us down to a detection limit of $\Mdot / \uten \sim
10^{-6} \Msunyr$. Figure 2 demonstrates that the limit is rather sensitive to
how soon after explosion the supernova is observed.  Earlier than 3
days is probably observationally unrealistic.

\section{Comparison with radio and X-ray limits} 

Radio limits have been presented for SN 1981B (Boffi \& Branch 1995) and SN
1986G (Eck et al.\ 1995), using scaling arguments. In particular, for
the close ($\sim$ 4 Mpc) SN 1986G, the range $10^{-7} - 3\times10^{-6}
\Msunyr$ was excluded. This limit is at the same sensitivity level as
we can expect from early high-resolution optical observations (cf. \S
2). However, as we noted in CLSPK96, radio limits are subject to large
systematic errors, because we do not know how efficiently the
synchrotron radiation is generated.  X-ray limits are potentially
firmer and more sensitive than both radio and optical limits.
Schlegel \& Petre (1993) estimated $\Mdot \la (2-3)\times10^{-6}
\Msunyr$ from {\sl ROSAT} observations of SN 1992A (distance $\sim$17
Mpc) at $\sim$16 days after explosion.  This is about an order of
magnitude more sensitive than we can expect from H$\alpha$ using the
estimate in \S 2.  However, Schlegel \& Petre did not consider X-ray
absorption in the CSM.  In Figure 3 we show the photon energy below
which the optical depth through the CSM is greater than unity (the
cutoff energy, $\epsilon_{\rm cut}$) for the faster model in Figure 1.
For the mass loss rates claimed to be excluded for SN 1992A, X-ray
absorption below $\sim 1$~keV is severe.  Schlegel \& Petre (1993)
observed in the range 0.2$-$2.4 keV, with a sensitivity peak around
$\sim 1$ keV, so their limit on the mass loss rate was probably too
optimistic.  Furthermore, for $\Mdot$ above $10^{-5} \Msunyr$, the
collapsed shocked ejecta probably block out all the flux in the
{\it ROSAT} range (cf.\ SN 1993J at early times; Fransson, Lundqvist,
\& Chevalier 1996), unless the CSM is aspherical and the interaction
region is viewed through the less dense part of the CSM.  More
reliable X-ray limits can be expected for photon energies in the 
range 5$-$10~keV.  

Searches for circumstellar matter are beginning to provide interesting
observational limits on the progenitor systems of SNe Ia.  The most
informative limits will come from X-ray telescopes like {\it AXAF}.
High-resolution optical spectroscopy is less sensitive, but has the
potential to provide information about velocities and abundances of
the CSM.  Radio observations are difficult to interpret.  At all
wavebands, observation as soon as possible after the explosion is
highly desirable.
\begin{figure}
\vspace{10cm}
\includegraphics{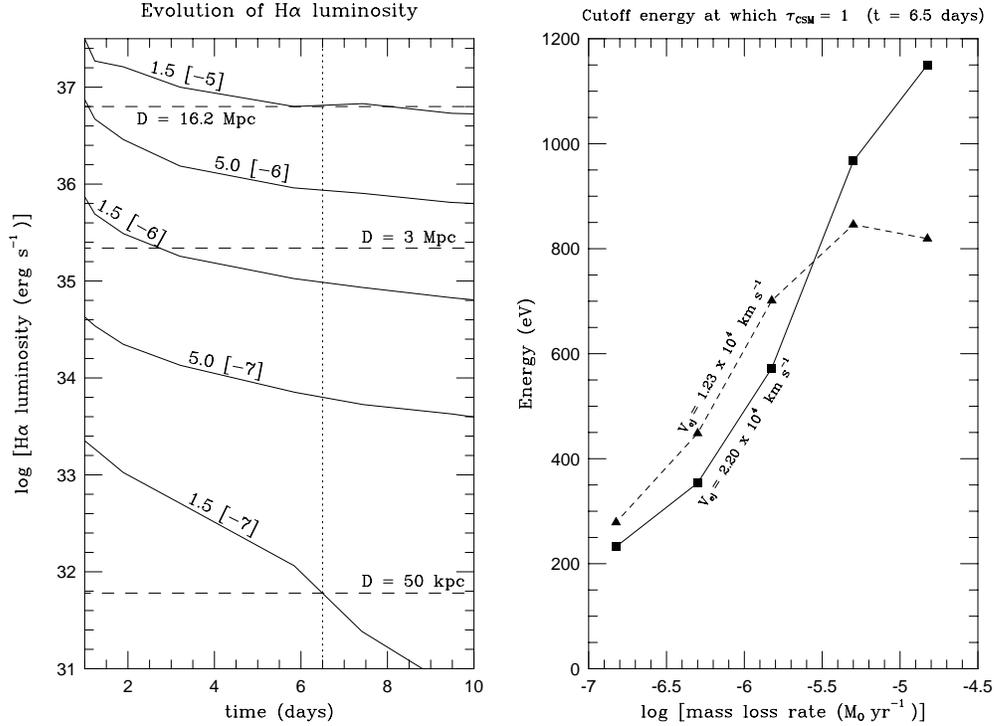}
\caption[]{(left) Evolution of H$\alpha$ luminosity for the models in
Figure 1.  Models are labeled with their $\Mdot / \uten$. The
vertical dotted line marks 6.5 days after explosion. The horizontal
dashed lines mark the sensitivity of the SN 1994D observation of
CLSPK96 at different distances.
\hskip 12cm 
Figure 3. (right) Cutoff energy, $\epsilon_{\rm cut}$, at which the optical depth
through the CSM is unity for the models in Figure 1. Below $\epsilon_{\rm cut}$
the CSM is opaque to X-ray emission from the supernova.}
\end{figure}
\vskip 0.3cm
{\bf Acknowledgements:} We thank David King, Linda Smith and Max Pettini for 
enlightening collaboration, and the Swedish Natural Science Research Council for
support.

\label{lastpage}

\end{document}